# A low-temperature synthesis of strongly thermochromic W and Sr co-doped VO$_2$ films with a low transition temperature

Michal Kaufman, Jaroslav Vlček,[a)] Jiří Houška, Sadoon Farrukh, Stanislav Haviar, Radomír Čerstvý, and Tomáš Kozák

Department of Physics and NTIS-European Centre of Excellence, University of West Bohemia, Univerzitní 8, 30100 Plzeň, Czech Republic

[a)]Electronic mail: vlcek@kfy.zcu.cz

## ABSTRACT

The reversible semiconductor-to-metal transition of vanadium dioxide (VO$_2$) makes VO$_2$-based coatings a promising candidate for thermochromic smart windows, reducing the energy consumption of buildings. We report low-temperature (320 °C) depositions of thermochromic V$_{1-x-y}$W$_x$Sr$_y$O$_2$ films with a thickness of 71–73 nm onto 170–175 nm thick Y-stabilized ZrO$_2$ layers on a 1 mm thick conventional soda-lime glass. The developed deposition technique is based on reactive high-power impulse magnetron sputtering with a pulsed O$_2$ flow feedback control allowing us to prepare crystalline W and Sr co-doped VO$_2$ films of the required stoichiometry without any substrate bias or post-deposition annealing. The W doping of VO$_2$ decreases the transition temperature below 25 °C, while the Sr doping of VO$_2$ increases the integral luminous transmittance, $T_{lum}$, significantly due to widening of the visible-range optical band gap, which is consistent with lowering of the absorption coefficient of films. We present the discussion of the effect of the Sr content in the metal sublattice of VO$_2$ on the electronic and crystal structure of V$_{1-x-y}$W$_x$Sr$_y$O$_2$ films, and on their temperature-dependent optical and electrical properties. An optimized V$_{0.855}$W$_{0.018}$Sr$_{0.127}$O$_2$ film exhibits a high $T_{lum}$ = 56.8% and modulation of the solar energy transmittance $\Delta T_{sol}$ = 8.3%, which are 1.50 times and 1.28 times, respectively, higher compared with those of the V$_{0.984}$W$_{0.016}$O$_2$ film. The achieved results constitute an important step toward a low-temperature synthesis of large-area thermochromic VO$_2$-based coatings for future smart-window applications, as it is easy to further increase the $T_{lum}$ and $\Delta T_{sol}$ by >6% and >3%, respectively, using a 280 nm thick top SiO$_2$ antireflection layer.

## I. INTRODUCTION

Vanadium dioxide (VO$_2$) is a thermochromic material that exhibits a reversible phase transition from a low-temperature monoclinic VO$_2$ (M1) semiconducting phase to a high-temperature tetragonal VO$_2$ (R) metallic phase at a transition temperature, $T_{tr}$, of approximately 68 °C for the bulk material.[1] This phase transformation is accompanied by dramatic changes in the electrical, optical, thermal and magnetic properties. The transformation can be initiated not only by temperature change but also by several other external stimuli such as light, terahertz pulses, electric and magnetic fields, pressure, and mechanical stress and strain. These characteristics make VO$_2$ a promising candidate for a wide variety of potential technological applications such as switches, memories, photodetectors, actuators, camouflages, passive radiators, resonators, sensors, field-effect transistors (see reviews[2-6] and the works cited therein), terahertz devices,[5] brain-inspired neuromorphic devices,[5,6] smart thermal radiator devices for spacecraft,[6] spintronic devices,[7,8] and energy-saving smart windows with automatically varied solar energy transmittance (see reviews[9-11] and the works cited therein).

Magnetron sputter deposition with its versatility and the ease of scaling up to large substrate sizes is probably the most important preparation technique of thermochromic $VO_2$-based coatings, particularly for future smart-window applications.[10,12-15] Note that magnetron sputter sources are used very frequently not only in coated-glass production lines (e.g., for deposition of low-emissivity coatings) but also in large-scale roll-to-roll deposition devices[16,17] producing coatings on ultrathin (0.02–0.20 mm) flexible glass or polymer foils. Here, it should be mentioned that the application potential of the thermochromic $VO_2$-based coatings depends on the ability to achieve not only the $VO_2$ stoichiometry but also the crystallization of the thermochromic $VO_2$ phase under sufficiently industry-friendly process conditions, i.e., at a maximum substrate surface temperature, $T_s$, during the preparation (deposition and possible post-annealing) close to 300 °C or lower[10,12,13,18,19] and without any substrate bias voltage[20]. The decrease of the deposition temperature of thermochromic $VO_2$-based coatings to 300 °C is of key importance: (i) to facilitate their large-scale production by reducing the energy consumption, simplifying substrate heating and cooling procedures and minimizing problems with a temperature non-uniformity over large substrate surfaces, and (ii) to allow deposition of these coatings onto temperature-sensitive flexible substrates. Note that traditional sputter techniques to prepare $VO_2$ coatings usually require a deposition or post-annealing temperature higher than 450 °C.[10,19,21] Moreover, desirable and challenging lowering of $T_{tr}$ toward room temperature using doping of $VO_2$ by other elements (such as W) should be performed without any degradation of thermochromic properties of doped $VO_2$ films.[5,10,22]

Reactive high-power impulse magnetron sputtering (HiPIMS) is a promising scalable[17,23] deposition technique for a low-temperature (300–350 °C) preparation of thermochromic $VO_2$-based films. HiPIMS is characterized by highly ionized fluxes of particles with high fractions of ionized sputtered metal atoms onto the substrate and by enhanced energies and momenta of the ions bombarding the growing films, allowing one to achieve film densification and crystallinity at a low substrate temperature and without a substrate bias voltage.[20] Recently, low-temperature HiPIMS depositions of thermochromic undoped $VO_2$ films[18,19,24-26] and strongly thermochromic $ZrO_2/V_{0.982}W_{0.018}O_2/ZrO_2$ coatings[27] have been reported.

In our previous paper,[28] we presented the design and optical properties of strongly thermochromic three-layer $YSZ/V_{0.855}W_{0.018}Sr_{0.127}O_2/SiO_2$ coatings, where YSZ is Y-stabilized $ZrO_2$, which were deposited onto conventional soda-lime glass (SLG) at a low substrate surface temperature $T_s = 320$ °C and without any substrate bias voltage. A coating design utilizing second-order interference in two antireflection layers was used to maximize both the integral luminous transmittance, $T_{lum}$, and the modulation of the solar energy transmittance, $\Delta T_{sol}$. A compact crystalline structure of the bottom YSZ antireflection layer improves the $VO_2$ crystallinity, while the top $SiO_2$ antireflection layer provides also the mechanical and environmental protection for the $V_{0.855}W_{0.018}Sr_{0.127}O_2$ layer. The coatings fulfill strict requirements for large-scale implementation on building glass (see review[10] and the works cited therein): $T_s$ of ≈ 300 °C or lower, $T_{tr}$ of ≈ 20 °C, $T_{lum} > 60\%$ and $\Delta T_{sol} > 10\%$. They exhibited $T_{tr} = 22$ °C with $T_{lum} = 63.7\%$ (below $T_{tr}$) and 60.7% (above $T_{tr}$), and $\Delta T_{sol} = 11.2\%$ for a $V_{0.855}W_{0.018}Sr_{0.127}O_2$ layer thickness of 71 nm. Our most recent measurements, carried out 12 months after depositions of these three-layer coatings, proved that there are no changes in their characteristics. To the best of our knowledge, the simultaneous fulfillment of these requirements had not been previously reported in the literature. A major challenge is to achieve the high $T_{lum}$ and $\Delta T_{sol}$ at a relatively low $T_{tr}$ and $T_s$.[4,5,10,11,21]

In this paper, we present in detail the principles of the developed sputter deposition technique, which was characterized only briefly in Ref. 28. This technique, based on reactive HiPIMS with a pulsed $O_2$ flow feedback control, allows us to prepare crystalline W and Sr co-doped $VO_2$ of the required stoichiometry. The W doping of $VO_2$ decreases the $T_{tr}$ to room temperature, while the Sr doping of $VO_2$ increases the $T_{lum}$ significantly due to widening of the visible-range optical band gap, which is consistent with lowering of the extinction coefficient of films in the whole visible range. We present the discussion of the effect of the Sr content in the metal sublattice of $VO_2$ on the electronic and crystal structure, and the optical, electrical and thermochromic properties of $V_{1-x-y}W_xSr_yO_2$ films deposited onto YSZ layers on SLG substrates. In this regard, the paper forms an extension of our recent work.[28] The study is a part of an overall program conducted to develop a new effective magnetron sputter technique for a low-temperature compatible fabrication of high-performance durable thermochromic $VO_2$-based multilayer coatings for smart-window applications.

## II. EXPERIMENTAL DETAILS

### A. Film preparation

The films were deposited onto 170–175 nm thick, highly transparent, YSZ layers[28] on 1 mm thick SLG substrates (25 × 75 mm$^2$), and onto Si(100) substrates (15 × 15 mm$^2$) in argon-oxygen gas mixtures at the argon partial pressure $p_{Ar}$ = 1 Pa, corresponding to the argon flow rate of 60 sccm. We used an ultra-high vacuum sputter device (ATC 2200-V AJA International Inc.) equipped by four unbalanced magnetrons with planar targets (diameter of 50 mm and thickness of 6 mm in all cases), located symmetrically around the vacuum chamber axis. The vacuum chamber (diameter of 560 mm and length of 430 mm) was evacuated by a turbomolecular pump (1.2 m$^3$s$^{-1}$) backed up with a double stage Roots pump (27 m$^3$h$^{-1}$). The base pressure before deposition was below 10$^{-4}$ Pa. The rotating (20 rpm) substrates at the distance of 145 mm from the targets were at a floating potential. The substrate surface temperature was 320 °C. This $T_s$ value was maintained during the deposition by a built-in infrared heating system calibrated using a thermocouple directly attached to the substrate surface (additional heating effect by the plasma included). The values of $p_{Ar}$ and of $p_{Ar} + p_{O_2}$, where $p_{O_2}$ is the oxygen partial pressure in the chamber, were measured at the chamber wall using a high-stability capacitance manometer (Baratron, Type 127, MKS) with an accuracy much better than 1%.

The $V_{1-x-y}W_xSr_yO_2$ films were deposited by controlled HiPIMS of a single V-W (4.0 wt.% corresponding to 1.14 at.%) target (99.95% purity) combined with a simultaneous pulsed DC magnetron sputtering of a Sr target (99.8% purity). The magnetron with the V-W target was driven by a unipolar high-power pulsed DC power supply (TruPlasma Highpulse 4002 TRUMPF Huettinger). The voltage pulse duration was 80 μs at a repetition frequency of 500 Hz (duty cycle of 4%) and the deposition-averaged target power density (spatially averaged over the total target area $A_t$ = 19.63 cm$^2$ in our case) was 14.9 Wcm$^{-2}$. The magnetron with the Sr target was driven by a unipolar pulsed DC power supply (IAP-1010 EN Technologies Inc.). To minimize arcing on the Sr target surface at an increased Sr target power density and to control a Sr content in the films easily, we used short 7 μs voltage pulses at a relatively high repetition frequency of 50 kHz (duty cycle of 35%) during the depositions with a pre-selected deposition-averaged target power density in the range from 0.1 Wcm$^{-2}$ to 3.3 Wcm$^{-2}$.

Oxygen was admitted into the vacuum chamber via a mass flow controller and two corundum conduits. Two O$_2$ inlets with a diameter of 1 mm were placed symmetrically above the V-W target racetrack at the same distance of 20 mm from the V-W target surface and oriented to the substrate. The to-substrate O$_2$ injection into the dense plasma in front of the sputtered target is very important for reactive HiPIMS depositions of oxide films.[29] It leads to a substantially decreased compound (oxide) fraction in the target surface layer, resulting in minimized arcing, increased sputtering of metal atoms, and low production of O$^-$ ions at the target, and to a substantially increased compound fraction in the substrate layer due to significantly increased chemisorption flux of O atoms and O$^+$ ions onto the substrate. This is caused by a considerably (2-3 times, as measured in Ref. 29) increased local oxygen partial pressure in front of the O$_2$ inlets, compared with a very low $p_{O_2}$ measured at the chamber wall, and by a very high degree of dissociation of the to-substrate injected O$_2$ molecules in the high-density plasma in front of the target. The total oxygen flow rate, $\Phi_{O_2}$, in both conduits was not fixed but alternating between two constant values $C_1$ (being in the range from 1.5 sccm to 2.0 sccm) and $C_2$ (being in the range from 2.5 sccm to 3.0 sccm) during a deposition (see Fig.1).

The moments of switching of the $\Phi_{O_2}$ values between $C_1$ and $C_2$ were determined during the deposition by a programmable logic controller using a pre-selected critical value of the average discharge current on the V-W target in a period of its power supply, $(\bar{I}_d)_{cr}$, being in the range from 0.55 A to 0.65 A: when $\bar{I}_d(t) < (\bar{I}_d)_{cr}$, $\Phi_{O_2} = C_2$ and when $\bar{I}_d(t) \geq (\bar{I}_d)_{cr}$, $\Phi_{O_2} = C_1$.

The used deposition technique allowed us to increase continuously the Sr content in the $V_{1-x-y}W_xSr_yO_2$ films and, as a result, to increase their $T_{lum}$, at a low substrate surface temperature $T_s = 320$ °C and a transition temperature $T_{tr} < 25$ °C. As can be seen, this cannot be achieved by only a simple increase in the Sr target power, but it requires a careful tuning of the process parameters $C_1$, $C_2$ and $(\bar{I}_d)_{cr}$.

The $V_{0.984}W_{0.016}O_2$ film was deposited by controlled HiPIMS of a V target (99.9% purity) combined with a simultaneous pulsed DC magnetron sputtering of a W target (99.9% purity). At the pre-selected $(\bar{I}_d)_{cr} = 0.46$ A, the $\Phi_{O_2}$ switched between 1.5 sccm and 1.9 sccm during the deposition performed using the same power supplies. For the V target, the voltage pulse duration was 80 μs at a repetition frequency of 500 Hz and the deposition-averaged target power density was 14.2 Wcm$^{-2}$. For the W target, the voltage pulse duration was 16 μs at a repetition frequency of 5 kHz (duty cycle of 8%) and the deposition-averaged target power density was 25 mWcm$^{-2}$.

## B. Film characterization

The W content in the metal sublattice of $V_{0.984}W_{0.016}O_2$, i.e., 1.6 ± 0.3 at.% of W, and the W and Sr contents in the metal sublattice of $V_{0.855}W_{0.018}Sr_{0.127}O_2$, i.e., 1.8 ± 0.2 at.% of W and 12.7 ± 1.8 at.% of Sr, and of $V_{0.740}W_{0.016}Sr_{0.244}O_2$, i.e., 1.6 ± 0.3 at.% of W and 24.4 ± 2.7 at.% of Sr, were measured in dedicated 440 nm thick films on Si(100) substrate in a scanning electron microscope (SU-70, Hitachi) using wave-dispersive spectroscopy (Magnaray, Thermo Scientific) at a low primary electron energy of 7.5 keV. Standard reference samples of pure V, W, $Fe_2O_3$ and $SrSO_4$ (Astimex Scientific Ltd.) were utilized. The room-temperature (25 °C) crystal structure of coatings was characterized by glancing incidence X-ray diffraction using a PANalytical X'Pert PRO diffractometer equipped with a Goebel mirror and working with a CuKα (40 kV, 40 mA) radiation at a glancing incidence of 0.6°.

The thickness and optical constants (refractive index, $n$, and extinction coefficient, $k$) of individual layers were measured by spectroscopic ellipsometry using the J.A. Woollam Co. Inc. VASE instrument equipped with an Instec heat/cool stage. The measurements of $n(\lambda)$ and $k(\lambda)$ were performed in the wavelength range of 300–2000 nm at the angles of incidence of 55°, 60° and 65° in reflection for $T_{ms} = -20$ °C (semiconducting state below $T_{tr}$) and $T_{mm} = 70$ °C (metallic state above $T_{tr}$). YSZ was described by the Cauchy dispersion formula, and $V_{0.984}W_{0.016}O_2$, $V_{0.855}W_{0.018}Sr_{0.127}O_2$ and $V_{0.740}W_{0.016}Sr_{0.244}O_2$ were represented by a combination of the Cody-Lorentz oscillator, Lorentz oscillators and (in the case of metallic phase) a Drude oscillator. The coating transmittance, $T$, and reflectance, $R$, were measured by spectrophotometry using the Agilent CARY 7000 instrument with an in-house made heat/cool cell. The measurements were performed in the wavelength range of 300–2500 nm at the angles of incidence of 0° ($T$) and 7° ($R$) for $T_{ms} = -20$ °C and $T_{mm} = 70$ °C. Hysteresis curves were measured for $T$ at $\lambda = 2500$ nm, $T_{2500}$, in the temperature range $T_m = -20$ °C to 70 °C. The coating performance is quantified by means of integral luminous transmittance, $T_{lum}(T_m)$, integral solar energy transmittance, $T_{sol}(T_m)$, and their modulations $\Delta T_{lum}$ and $\Delta T_{sol}$. These quantities are defined as

$$T_{\text{lum}}(T_{\text{m}}) = \frac{\int_{380}^{780} \varphi_{\text{lum}}(\lambda)\varphi_{\text{sol}}(\lambda)T(T_{\text{m}},\lambda)d\lambda}{\int_{380}^{780} \varphi_{\text{lum}}(\lambda)\varphi_{\text{sol}}(\lambda)d\lambda},$$

$$\Delta T_{\text{lum}} = T_{\text{lum}}(T_{\text{ms}}) - T_{\text{lum}}(T_{\text{mm}}),$$

$$T_{\text{sol}}(T_{\text{m}}) = \frac{\int_{300}^{2500} \varphi_{\text{sol}}(\lambda)T(T_{\text{m}},\lambda)d\lambda}{\int_{300}^{2500} \varphi_{\text{sol}}(\lambda)d\lambda},$$

$$\Delta T_{\text{sol}} = T_{\text{sol}}(T_{\text{ms}}) - T_{\text{sol}}(T_{\text{mm}}),$$

where $\varphi_{\text{lum}}$ is the luminous sensitivity of the human eye and $\varphi_{\text{sol}}$ is the solar irradiance spectrum at an air mass of 1.5.[30]

The optical band gaps $E_{g1}$ and $E_{g2}$ were determined from Tauc plots by utilizing the relation $(\alpha E)^{1/2} \sim E - E_g$, valid for indirect-allowed optical transitions,[31-33] where $E$ is the photon energy and $\alpha$ is the absorption coefficient (neglecting the absorption in glass and YSZ) calculated using the Beer-Lambert law as $\alpha = -[\ln(T/1-R)]/d$,[31,33] where $d$ is the thickness of the thermochromic film.

The temperature-dependent electrical properties (electrical resistivity, $\rho$, and the Hall coefficient, $R_h$) were measured in the Van der Pauw configuration using a Variable Temperature Hall Measurement System (MMR Technologies) equipped with a Joule-Thomson refrigerator and a heating stage. Gold contacts (thickness of 50 nm) were sputter-deposited in the corners of the 9 × 9 mm² samples to improve the contact between the sample and the gold spring-loaded probe tips. Electrical resistivity, $\rho$, was measured from $T_m$ = -20 °C to 70 °C with the step of 10 °C. The Hall coefficient, $R_h$, was measured for selected temperatures using a magnetic field strength of 1.4 T and the majority carrier density, $n_{\text{carriers}}$, was calculated as $1/(eR_h)$.

## III. RESULTS AND DISCUSSION

In this section, we present the principles of the developed sputter deposition technique and the results achieved for three thermochromic films with a thickness of 71–73 nm deposited onto 170–175 nm thick YSZ layers on 1 mm thick SLG substrates: an optimized $V_{0.855}W_{0.018}Sr_{0.127}O_2$ film with a high $T_{\text{lum}}(T_{\text{mm}}) = 55.6\%$ for the high-temperature state and the highest achieved $\Delta T_{\text{sol}} = 8.3\%$, a $V_{0.740}W_{0.016}Sr_{0.244}O_2$ film with an even higher $T_{\text{lum}}(T_{\text{mm}}) = 64.3\%$ but at a decreased $\Delta T_{\text{sol}} = 5.0\%$ (a further increase in the Sr content resulted in a drop of $\Delta T_{\text{sol}}$ to zero), and an optimized $V_{0.984}W_{0.016}O_2$ film (prepared for comparative purposes without Sr) with $T_{\text{lum}}(T_{\text{mm}}) = 37.3\%$ and $\Delta T_{\text{sol}} = 6.5\%$ (see Table I). The presented $V_{1-x-y}W_xSr_yO_2$ films were selected on the basis of the measured $T_{\text{lum}}$, $\Delta T_{sol}$ and $T_{\text{tr}}$ values, supported by X-ray diffraction measurements, from several tens of films prepared under systematically changed deposition conditions. The films exhibited good uniformity and reproducibility of the structure and properties. For example, a maximum absolute deviation of 0.3% and 0.4% in the values of $T_{\text{lum}}$ and $\Delta T_{\text{sol}}$, respectively, was measured at the same $T_{\text{tr}}$ = 24 °C on 6 individual smaller (10 × 25 mm²) samples cut from an original larger sample of $V_{0.855}W_{0.018}Sr_{0.127}O_2$ after deposition. No changes in their X-ray diffraction patterns were observed.

## A. Discharge characteristics

The principles of the developed sputter deposition technique are illustrated in Fig.1 for the deposition of the optimized thermochromic $V_{0.855}W_{0.018}Sr_{0.127}O_2$ film. The time evolutions of the magnetron voltage, $U_d(t)$, and the target current density $J_t(t) = I_d(t)/A_t$, where $I_d(t)$ is the discharge current, are shown for both targets at the minimum and maximum value of the oxygen partial pressure $p_{O_2}$ = 26 mPa and 55 mPa, respectively, in the vacuum chamber corresponding to the minimum and maximum value of the average discharge current on the V-W target in a period of its power supply, $\bar{I}_d$, respectively, during the deposition.

The pulsed $O_2$ flow feedback control with a to-substrate $O_2$ injection into the dense plasma in front of the sputtered V-W target makes it possible to deliver a high power into discharge pulses with minimized arcing on the V-W target surface and thus, to utilize the exclusive benefits of the HiPIMS discharges (such as highly ionized fluxes of particles with high fractions of ionized sputtered V atoms onto the substrate and with enhanced energies and momenta of the ions bombarding the growing films,[20] and a high chemisorption flux of O atoms and $O^+$ ions onto the substrate at a low oxygen partial pressure in the vacuum chamber mainly due to the used to-substrate $O_2$ injection into the dense plasma[29]) in the preparation of W and Sr co-doped $VO_2$ films with a highly crystalline thermochromic $VO_2$ phase on unbiased substrates at a low $T_s$ = 320 °C. The advantages of the used pulsed $O_2$ flow control are: (i) very high stability of the deposition process, as this process control has no problems with a usual delay of valves and inertia of the inlet system caused by the transfer of oxygen gas through the conduits (see the inset of Fig. 1), and (ii) simplicity, as no additional measurement devices (such as a plasma monitoring system, mass spectrometer or lambda sensor) are needed.

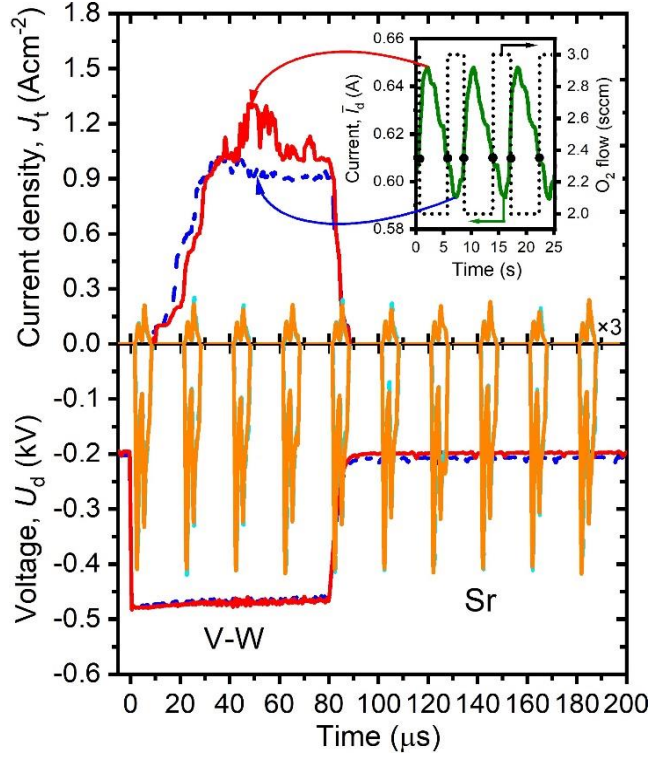

FIG. 1. Waveforms of the magnetron voltage, $U_d$, and of the target current density, $J_t$, for a preset deposition-averaged target power density of 14.9 Wcm$^{-2}$ for a composite V-W target and 1.9 Wcm$^{-2}$ for a Sr target, respectively, at the minimum $p_{O_2}$ = 26 mPa (blue and turquoise lines) and the maximum $p_{O_2}$ = 55 mPa (red and orange lines) during a deposition of the optimized thermochromic $V_{0.855}W_{0.018}Sr_{0.127}O_2$ films (the $J_t$ values for Sr target are magnified 3 times). Time evolution of the average discharge current on the V-W target in a period of its power supply, $\bar{I}_d$, during the deposition is shown in the inset. A pre-selected critical value $(\bar{I}_d)_{cr}$ = 0.61 A determining the switching between the oxygen flow rates $\Phi_{O_2}$ = 2.0 sccm and $\Phi_{O_2}$ = 3.0 sccm is marked by dots.

**B. Electronic structure and optical properties of films**

Figure 2a shows a schematic energy band diagram for the low-temperature VO$_2$(M1) semiconducting phase with two band gaps. $E_{g2}$ is a band gap between the filled lower part of the split $d_\parallel$ band and the empty $\pi^*$ band while $E_{g1}$ is a band gap between the filled $\pi$ band and the empty $\pi^*$ band. The width of the $E_{g2}$ gap is in the infrared spectral range with reported values of about 0.6 eV.[36-38] At the transition to the high-temperature VO$_2$(R) metallic phase, $E_{g2}$ closes. This results in an increased density of free charge carriers with a direct effect mainly on the contribution of infrared wavelengths to $\Delta T_{sol}$. The width of the $E_{g1}$ gap is in the visible range with reported values of 1.5–1.7 eV.[31,32,39,40] Considering the associated interband transitions[37,38] and their direct effect on $T_{lum}$ and the coating color, it makes understandable a worldwide effort to improve these characteristics via controlling $E_{g1}$.

As can be seen in Fig. 2b, the incorporation of 12.7 at.% and 24.4 at.% of Sr into the metal sublattice of VO$_2$ at an almost fixed W content of 1.6–1.8 at.% led to a decrease of $E_{g2}$ from 0.53 eV to 0.39 eV and 0.28 eV, respectively, and to a substantial widening of the visible-range gap with an increase in $E_{g1}$ from 1.51 eV to 1.74 eV and 2.23 eV, respectively, for the low-temperature state ($T_{ms}$ = -20 °C) of the doped VO$_2$ films. This takes place in parallel to the qualitatively preserved thermochromic effect: the $\alpha$ values measured for the high-temperature state ($T_{mm}$ = 70 °C) of each film in the infrared range are considerably higher than those for the low-temperature state ($T_{ms}$ = -20 °C).

Qualitatively, the direction and strength of the role of Sr is in accordance with what its basic characteristics promise. The bonding in understoichiometric VO$_2$ cannot be purely ionic due to the one excess electron per formula unit, but is in part covalent/metallic. This constitutes the usual recipe for narrow band gap metal oxides: Cr$_2$O$_3$, Fe$_2$O$_3$, MnO$_2$ or PbO. The partial replacement of V (5 valence electrons) by Sr (2 valence electrons) decreases the electronic excess and makes the bonding more ionic. The ionic character of the bonding is further supported by the lower Pauling electronegativity of Sr (0.95) compared to V (1.63), and constitutes the usual recipe for opening the band gap: note the wide band gap (>6 eV) of not only SrO but also other alkaline earth metal oxides CaO, MgO or BeO.

Quantitatively, the gradient characterizing the enhancement of $E_{g1}$ of V$_{0.855}$W$_{0.018}$Sr$_{0.127}$O$_2$, (1.74 - 1.51)/12.7 = 18 meV/at.% of Sr, is in a very good agreement with that reported at about the same doping level in Ref.39: V$_{0.991}$W$_{0.009}$O$_2$ with $E_{g1}$ = 1.73 eV and V$_{0.872}$W$_{0.009}$Sr$_{0.119}$O$_2$ with $E_{g1}$ = 1.94 eV leading to a gradient (1.94 - 1.73)/11.9 = also 18 meV/at.% of Sr. The data reported in Ref. 39 were obtained for ≈100 nm thick films deposited by RF magnetron sputtering at $T_s$ = 450 °C onto 50 nm thick SnO$_2$ layers on quartz glass substrates and using a more simplified formula for $\alpha$. Note that the calculated gradient constitutes a lower bound on its true value, which may be larger in a case (indicated by the electrical properties discussed in Sec. IIIC) of not incorporating all Sr into the VO$_2$ lattice.

The enhancement of $E_{g1}$ is consistent with lowering of the extinction coefficient (Fig. 3b) and in turn absorption coefficient (Fig. 2b) in the whole visible range. The lower values of $k$ due to the Sr incorporation open a possibility to prepare thermochromic films with a higher $T_{lum}$ (at a given thickness) or a higher $\Delta T_{sol}$ (because a lower $k$ at 550 nm allows a higher thickness) or both. In parallel, the Sr incorporation decreases the refractive index in the visible range and at low $T_{ms}$ also in the infrared range (Fig. 3a). This is, once again, in accordance with the basic atomic characteristics: the lower number of valence electrons of Sr (only partially compensated by its larger radius) decreases the polarizability per metal atom and even more the polarizability per unit volume, which is at low $k$ directly related to $n$. Note, yet once again, the low $n$ (<2) of not only SrO but also other alkaline earth metal oxides, such as CaO, MgO or BeO. The lower value of $n$ measured at 550 nm is of key importance for the selection of a suitable top antireflection layer.[28]

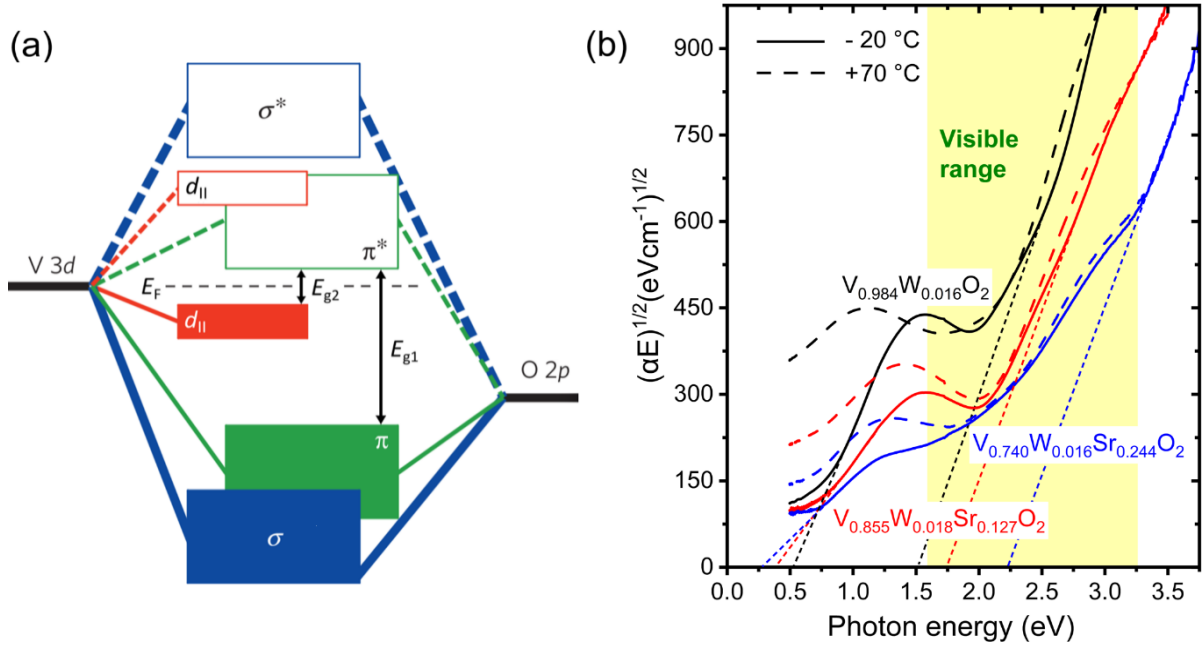

FIG. 2. (a) Schematic energy band diagram for pure $VO_2$(M1) semiconducting phase, based on Ref. 34, with two gaps $E_{g1}$ and $E_{g2}$ according to Goodenough[31,35]. $E_F$ denotes the Fermi energy. (b) $(\alpha E)^{1/2}$ as a function of the photon energy $E$, where $\alpha$ is the absorption coefficient, measured at $T_{ms}$ = -20 °C and $T_{mm}$ = 70 °C for three doped $VO_2$ films (71–73 nm) deposited onto YSZ layers (170–175 nm) on 1 mm thick SLG substrates. At -20 °C, linear fittings are performed to determine $E_{g1}$ and $E_{g2}$ (Tauc plots). The shaded area represents the visible range of the electromagnetic spectrum (380–780 nm).

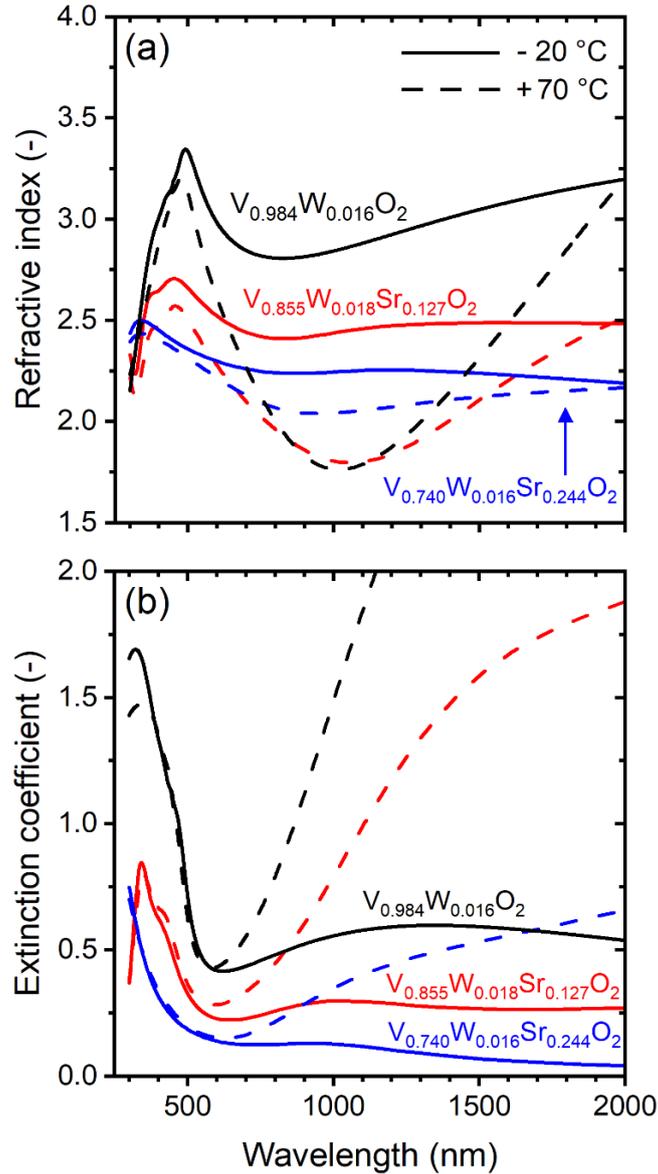

FIG. 3. Spectral dependences of the refractive index (a) and the extinction coefficient (b) measured at $T_{ms}$ = -20 °C and $T_{mm}$ = 70 °C for three doped $VO_2$ films (71–73 nm) deposited onto YSZ layers (170–175 nm) on 1 mm thick SLG substrates.

## C. Crystal structure and thermochromic behavior of films

Figure 4 shows the X-ray diffraction patterns taken at 25 °C from two investigated W and Sr co-doped $VO_2$ films prepared on YSZ layers. As can be seen, the YSZ layers are well crystalline, see the strong narrow peaks close to the positions reported for tetragonal $Y_{0.06}Zr_{0.94}O_{1.97}$ phase (PDF #04-021-9607)[41]. This confirms their role of a structure template improving crystallinity of the thermochromic $VO_2$-based films.[42] The $V_{0.855}W_{0.018}Sr_{0127}O_2$ film contributes by diffraction peaks very close to the positions of both $VO_2$(M1) (PDF #04-003-2035) and $VO_2$(R) (PDF #01-073-2362). These two desired phases are difficult to distinguish and they are actually expected to be present simultaneously because the measurement temperature $T_m$ = 25 °C is close to the transition temperature $T_{tr}$ = 22–24 °C (as shown later). The $V_{0.740}W_{0.016}Sr_{0.244}O_2$ film with a relatively high Sr content in the metal sublattice of $VO_2$ exhibits only very low diffraction peaks of the thermochromic $VO_2$ phases owing to their decreased crystallinity (smaller size of crystal grains with wider amorphous intergrain regions). The important piece of information in Figure 4 is the absence of any other peaks: there are no fingerprints of non-thermochromic (in the $T_m$ range of interest) stoichiometries, such as $V_2O_3$, $V_4O_9$ or $V_2O_5$, or polymorphs such as $VO_2(P)$ or $VO_2(B)$. This confirms the success of our controlled sputter deposition technique. Here, it should be mentioned that there exist tens of vanadium suboxides, among which $VO_2$ is not the most thermodynamically stable phase,[2,9,10] and more than ten $VO_2$ polymorphs[5,10].

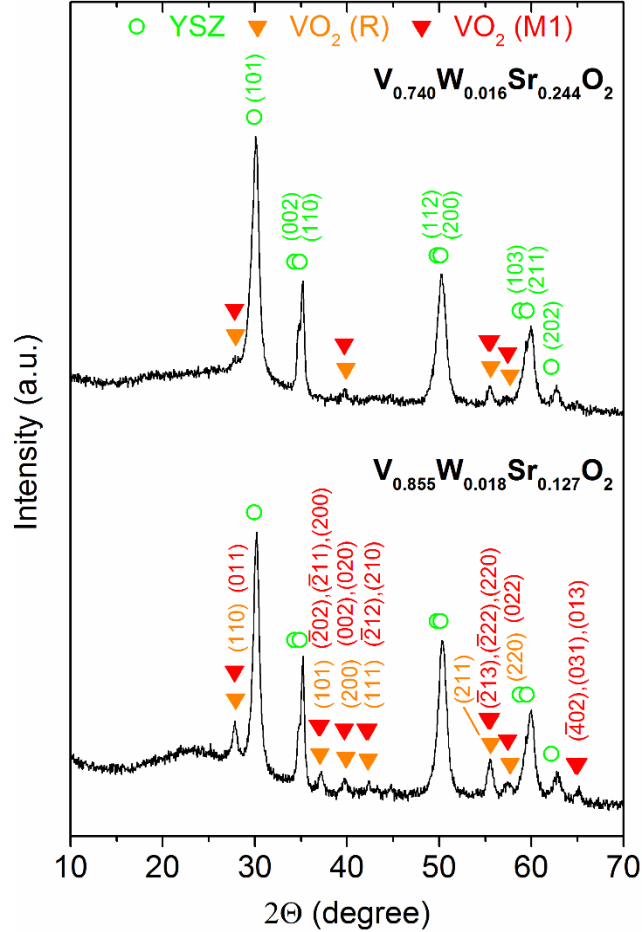

FIG. 4. X-ray diffraction patterns taken at $T_m = 25$ °C from two doped $VO_2$ films (71–72 nm) deposited onto YSZ layers (170–175 nm) on 1 mm thick SLG substrates. The main diffraction peaks of $VO_2(M1)$, $VO_2(R)$ and YSZ (tetragonal $Y_{0.06}Zr_{0.94}O_{1.97}$) are marked.

The thermochromic behavior of the three doped $VO_2$ films is presented in Figs. 5 and 6, and in Table I. Figure 5 captures the role of Sr, leading, in agreement with the presented widening of $E_{g1}$ in Fig. 2b and lowering of $k(\lambda)$ in Fig. 3b, to a significantly enhanced $T(\lambda)$ (at about the same thickness of the doped $VO_2$ films) at both measurement temperatures. While the transmittance enhancement is arguably most important in the visible, it takes place in the whole $\lambda$ range investigated. The spectral transmittance was used to calculate the integral transmittances $T_{lum}(T_m)$ and $T_{sol}(T_m)$, and their modulations, provided in Table I. As can be seen, the incorporation of 12.7 at.% and 24.4 at.% of Sr into the metal sublattice led to an increase in $T_{lum}(T_{mm})$ from 37.3% to 55.6% and 64.3%, respectively, for the high-temperature state, and in $T_{lum}(T_{ms})$ from 37.9% to 56.8% and 64.9%, respectively, for the low-temperature state. The incorporation of 12.7 at.% of Sr resulted in an increase in $\Delta T_{sol}$ from 6.5% to 8.3%, but $\Delta T_{sol}$ decreased to 5.0% for 24.4 at.% of Sr. The enhancement of $T_{lum}$ is due to the Sr-induced lowering of $k(\lambda)$ in the visible range, while the enhancement of $\Delta T_{sol}$ for 12.7 at.% of Sr is due to the contribution of both visible wavelengths (because the Sr-induced lowering of $k(\lambda)$ is larger in the low-temperature state, see the enhancement of $\Delta T_{lum}$ from 0.7% to 1.2%) and the shortest infrared wavelengths (because of the Sr-induced stronger modulation of $T(\lambda)$). The lower $\Delta T_{sol}$ for 24.4 at.% of Sr is caused by a decreased crystallinity of the thermochromic $VO_2$ phases (see Fig. 4). Let us emphasize that while all aforementioned data clarify the qualitative effect of Sr, they do not constitute the maximum achievable coating performance: previously[28] we have shown that a proper top antireflection layer further increases $T_{lum}$ and $\Delta T_{sol}$ by more than 6% and 3%, respectively.

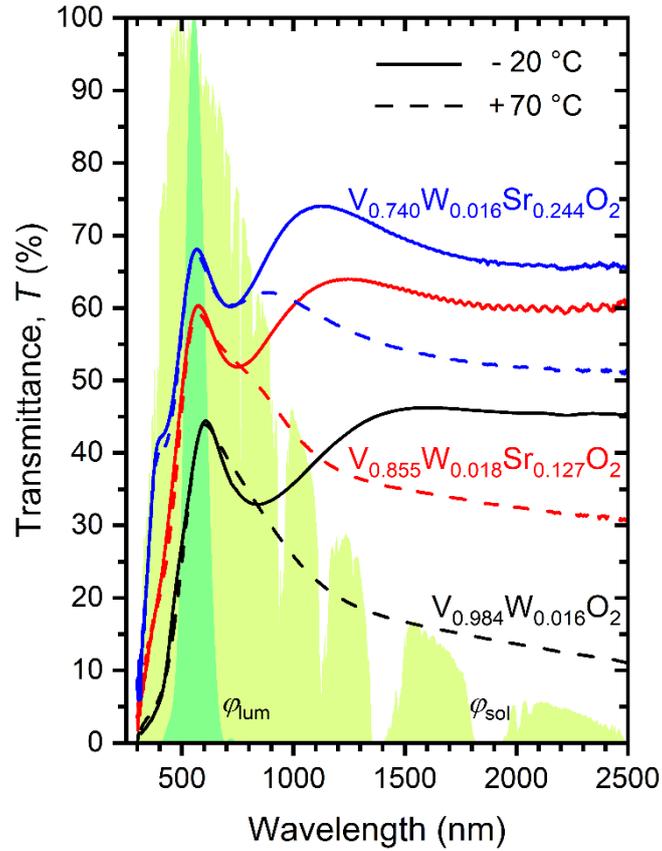

FIG. 5. Spectral transmittances measured at $T_{ms}$ = -20 °C and $T_{mm}$ = 70 °C for three doped $VO_2$ films (71–73 nm) deposited onto YSZ layers (170–175 nm) on 1 mm thick SLG substrates. The contours of the shaded areas represent the luminous sensitivity of the human eye ($\varphi_{lum}$) and the solar irradiance spectrum ($\varphi_{sol}$), normalized to maxima of 100%.

TABLE I. The integral luminous and solar energy transmittance ($T_{lum}(T_m)$ and $T_{sol}(T_m)$, respectively) measured at $T_{ms}$ = -20 °C and $T_{mm}$ = 70 °C, together with the corresponding modulations $\Delta T_{lum}$ and $\Delta T_{sol}$, and the transition temperature $T_{tr}$ (see Fig. 6) for three doped $VO_2$ films (71–73 nm) deposited onto YSZ layers (170–175 nm) on 1 mm thick SLG substrates. Note that a 280 nm thick top $SiO_2$ antireflection layer on these two-layer coatings increases the $T_{lum}$ and $\Delta T_{sol}$ by more than 6% and 3%, respectively.[28] Moreover, it provides also the mechanical and environmental protection for the thermochromic layers.

| Sample | $T_{lum}(T_{ms})$ (%) | $T_{lum}(T_{mm})$ (%) | $\Delta T_{lum}$ (%) | $T_{sol}(T_{ms})$ (%) | $T_{sol}(T_{mm})$ (%) | $\Delta T_{sol}$ (%) | $T_{tr}$ (°C) |
|---|---|---|---|---|---|---|---|
| $V_{0.984}W_{0.016}O_2$ | 37.9 | 37.3 | 0.6 | 33.9 | 27.4 | 6.5 | 25–26 |
| $V_{0.855}W_{0.018}Sr_{0.127}O_2$ | 56.8 | 55.6 | 1.2 | 52.4 | 44.1 | 8.3 | 22–24 |
| $V_{0.740}W_{0.016}Sr_{0.244}O_2$ | 64.9 | 64.3 | 0.6 | 62.2 | 57.2 | 5.0 | 23–24 |

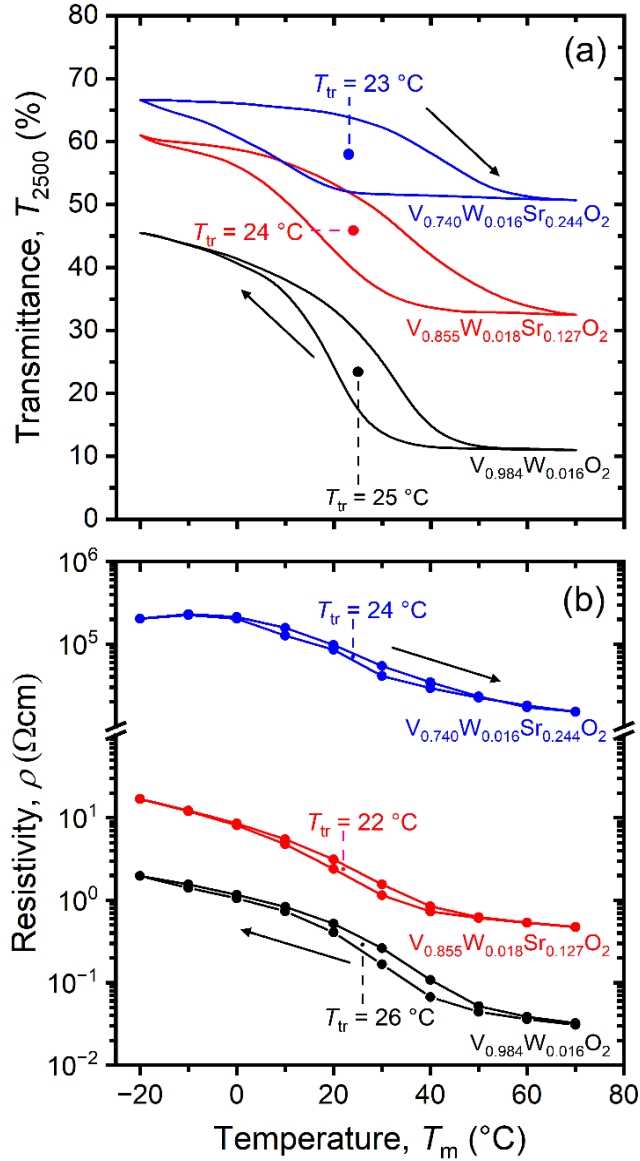

FIG. 6. Temperature dependences of the transmittance at 2500 nm (a) and the electrical resistivity (b) for three doped $VO_2$ films (71–73 nm) deposited onto YSZ layers (170–175 nm) on 1 mm thick SLG substrates. The transition temperatures are also presented.

Figure 6 examines the thermochromic transition of the three investigated doped $VO_2$ films in terms of the temperature dependence of $T_{2500}$ (Fig. 6a) and $\rho$ (Fig. 6b) between -20 °C and 70 °C. Both quantities yield almost the same transition temperatures. First, it can be seen that the doping of $VO_2$ with 1.6 at.% of W in the metal sublattice of $V_{0.984}W_{0.016}O_2$ films (that is, destabilization of the low-temperature semiconducting phase by the larger size and extra valence electron of W compared to V) allowed us to lower $T_{tr}$ = 56–57 °C achieved for pure $VO_2$ films[20] prepared using HiPIMS deposition in our laboratory to 25–26 °C. Second, this successful lowering of $T_{tr}$ is almost independent of the subsequent incorporation of Sr: an approximately fixed W content of 1.6–1.8 at.% led to an approximately fixed $T_{tr}$ = 22–26 °C despite the very wide range of Sr content between 0 and 24.4 at.%. A case can be made that the potential of Sr to (i) increase $T_{tr}$ due to less valence electrons and (ii) decrease $T_{tr}$ due to the unit cell expansion leading to weaker V-V bonds,[39] is either very low or almost compensating each other. Note that (contrary to some other results or even generalizing statements in the literature) the sputter deposition technique used allowed us to lower $T_{tr}$ at preserved strongly thermochromic behavior.

Third, increasing incorporation of Sr resulted in increasing $T_{2500}$ and $\rho$ (at a given $T_m$), and in weaker dependencies of $T_{2500}$ and $\rho$ on $T_m$. Quantitatively, moving from $V_{0.984}W_{0.016}O_2$ through the optimized $V_{0.855}W_{0.018}Sr_{0.127}O_2$ to $V_{0.740}W_{0.016}Sr_{0.244}O_2$ shifts the interval of the measured $T_{2500}(T_m)$ values from 11–46% through 32–61% to 51–67% and the interval of the measured $\rho(T_m)$ values from 0.03–2.0 Ωcm (at $n_{carriers}$ of $2.7\times10^{21}$–$1.8\times10^{19}$ cm$^{-3}$) through 0.47–17.0 Ωcm (at $n_{carriers}$ of $5\times10^{19}$–$\approx10^{18}$ cm$^{-3}$) to $1.5\times10^4$–$2.0\times10^5$ Ωcm (at unmeasurably low $n_{carriers}$) with a decreasing modulation of $\rho$ from 67 times through 36 times to 13 times, respectively. Interestingly, the material $V_{0.984}W_{0.016}O_2$ clearly exhibited a p-type conductivity: arguably given more by the characteristics of $VO_2$ than by its n-doping by W (this is further supported by the fact that $\rho$ of doped $V_{0.984}W_{0.016}O_2$ is in fact higher than previously[20] measured $\rho$ of undoped $VO_2$).

Note that the effect of Sr on $\rho$ is larger than that of $T_m$, and that its direction correlates with the wide band gap of pure SrO rather than with $E_{g2}$ of $V_{1-x-y}W_xSr_yO_2$ (let alone that the latter is relevant only for the low-temperature phase). Specifically, the (formally speaking) metallic high-temperature state of $V_{0.740}W_{0.016}Sr_{0.244}O_2$ is actually characterized by orders of magnitude higher $\rho$ and lower $n_{carriers}$ than the semiconducting low-temperature state of $V_{0.984}W_{0.016}O_2$ and $V_{0.855}W_{0.018}Sr_{0.127}O_2$. This strongly indicates that at 24.4 at.% of Sr (and to a much smaller extent perhaps already at 12.7 at.% of Sr) part of Sr is not embedded in the metal sublattice of $VO_2$ crystals but in an amorphous (not observable by XRD in Fig. 4) matrix. Thus, the high $\rho$ values of Sr-rich $V_{1-x-y}W_xSr_yO_2$ should be understood as $\rho$ of nanocomposites formed by $VO_2$-based crystals embedded in a high-resistivity SrO-based matrix.

The larger width of the middle section of the hysteresis loop of $T_{2500}$ (Fig. 6a) compared to that of $\rho$ (Fig. 6b) may be explained by larger hysteresis width of smaller crystals.[43,44] On the one hand, the transmittance is given by characteristics of all crystals including the relatively small ones, leading to a considerable hysteresis width. On the other hand, a case can be made that most of the electrical conduction takes place through channels which consist of relatively few relatively large crystals, decreasing the hysteresis width.

## IV. CONCLUSIONS

We have presented and explained a scalable industry-friendly technique for sputter deposition of strongly thermochromic $V_{1-x-y}W_xSr_yO_2$ films with a transition temperature $T_{tr} < 25$ °C. The combination of (i) exclusive benefits of HiPIMS of a V-W target, (ii) mid-frequency pulsed DC magnetron sputtering of a Sr target and (iii) pulsed $O_2$ flow feedback control with a proper geometry of $O_2$ injection allowed us to prepare highly crystalline doped $VO_2$-based films on unbiased substrates at low temperature (320 °C) and without any tradeoffs resulting from lowering $T_{tr}$ by W.

Next, as the deposition technique developed allowed us to smoothly increase the Sr content in the films at minimized arcing on any target surface, we have studied the effect of Sr content on the performance of $YSZ/V_{1-x-y}W_xSr_yO_2$ bilayers at approximately fixed thicknesses. First, there is a monotonic effect of Sr on opening the visible-range band gap, lowering the extinction coefficient, increasing the luminous transmittance and increasing the resistivity. Second, there is a non-monotonic effect of Sr on the modulation of solar energy transmittance leading to the existence of an optimum Sr content. Third, there is only a very weak effect of Sr on $T_{tr}$.

Taking into account that (i) it is possible to define an appropriate elemental composition of single-compound V-W-Sr magnetron targets for large-scale deposition devices now, and (ii) it is easy to increase substantially the luminous transmittance and the modulation of solar energy transmittance of the $YSZ/V_{1-x-y}W_xSr_yO_2$ bilayers using a top $SiO_2$ antireflection layer, which provides also the mechanical and environmental protection for the thermochromic layer, the results constitute an important step toward large-scale energy-saving applications of smart windows covered by thermochromic coatings with high performance.

A desirable further reduction in the deposition temperature might be achieved by optimization of the magnetron sputter technique, by its combination with a subsequent flash lamp annealing of layers and by using a crystalline template (serving also as a highly optically transparent bottom antireflection layer, see YSZ in our case) which would ensure epitaxial growth of $VO_2$-based layer. The epitaxial growth of the $VO_2$-based layer could result in an increased modulation of solar energy transmittance, a sharper thermochromic transition and its narrower hysteresis.

## ACKNOWLEDGMENTS

This work was supported by the project Quantum materials for applications in sustainable technologies (QM4ST), funded as Project No. CZ.02.01.01/00/22_008/0004572 by Programme Johannes Amos Comenius, call Excellent Research.

## AUTHOR DECLARATIONS

**Conflict of Interest**

The authors have no conflict to disclose.

## DATA AVAILABILITY

The data that support the findings of this study are available from the corresponding author upon reasonable request.